\documentstyle[aps,twocolumn]{revtex}
\begin{document}
\draft
\title{Mass-Shell Behavior of Electron  Propagator at Low Temperature}
\author{H. Arthur Weldon}
\address{Department of Physics,
West Virginia University, Morgantown, West Virginia, 26506-6315}
\date{October 11, 1998}
\maketitle
\begin{abstract}
At $T=0$ the full electron propagator is known to have an infrared
anomalous dimension and thus  a branch point at 
 $P^{2}=m^{2}$ rather than a  pole. 
An explicit calculation shows that 
if  $0<eT\ll m$ the retarded
self-energy is analytic in the vicinity of $P^{2}\approx m^{2}$, which includes 
the thermal mass-shell. The low-temperature propagator has a simple pole. Only  
when  $T=0$ is there a branch point at the mass-shell. 
\end{abstract}


\section{Introduction}

In high temperature gauge theories the fermion propagator is quite
different than at zero temperature \cite{propagator} and calculations require
the  Braaten-Pisarski resummation of hard thermal loops \cite{BP1,BP2,Rebhan}.  One
of the important quantities to be calculated is the imaginary part of the fermion
self-energy, or damping rate, which has been computed in various circumstances. In
QCD the massless gluons  are so changed by thermal effects that 
 a resummed gluon propagator is always necessary to compute thermal damping rates.
Whether the quark masses are small  or large compared to $gT$ determines
if a resummed quark propagator is required. The quark damping rate has been
computed in both cases \cite{BP1,quark}
and the potential infrared divergences  are controlled by  incorporating a
magnetic screening mass.
The  absence of a magnetic screening mass in QED makes the electromagnetic
damping rates more problematic \cite{QED}.
It appears that at high temperature ($eT\gg m$) the electron propagator at large
time does not decay exponentially 
\cite{Blaizot}. The same behavior is observed numerically in scalar
QED \cite{Boyanovsky}.

In QED at low temperature ($eT\ll m$) neither the electron nor photon propagator
require Braaten-Pisarski resummation. One would expect very little qualitative
difference between low temperature and zero temperature. However, explicit
calculation will show that there is an important difference: at $T=0$ the electron
propagator has a branch cut at the mass shell but for  $T>0$ the propagator
has a simple pole.

\subsection{Zero Temperature}

The scattering formalism of zero-temperature quantum field theory relies upon the
assumption that asymptotically separated particles do not influence each other.
Consequently propagators are supposed to have simple poles at the physical mass of
the particle. However this argument fails for charged particles 
because of the long range of the Coulomb force.  Two charged particles that are
arbitrarily far apart do not travel in straight lines. Instead their asymptotic
trajectories are bent and the curvature  grows logarithmically with their
separation. Thus charged particles that are infinitely separated cannot be treated
as free particles.
The consequences of this for  QED were first investigated by
Abrikosov, Chung, and Kibble \cite{Abrikosov,Chung,Kibble}. They   showed that
the electron propagator actually has a branch point at $P^{2}=m^{2}$ rather than a
pole. The propagator has the behavior 
\begin{equation}S^{\prime}(P)\to {Z_{2}\over m^{2\gamma}}{\not\!P
+m\over
(P^{2}-m^{2})^{1-\gamma}}\end{equation}
in the vicinity $P^{2}\approx m^{2}$. 
The value of $\gamma$
depends upon the choice of gauge. If the free photon propagator is
\begin{displaymath}D^{\mu\nu}(K)=-{g^{\mu\nu}\over K^{2}}+(1-\xi)
{K^{\mu}K^{\nu}
\over (K^{2})^{2}}\end{displaymath}
then 
\begin{equation}
\gamma=(\xi-3)\alpha/2\pi.\label{gamma}\end{equation}
  Only when $\xi =3$ (Yennie gauge)  does the
electron propagator have a simple pole. The exponent $\gamma$ is the infrared
anomalous dimension of the electron propagator. A recent calculation [4] of
$\gamma$ for the propagator and other multi-fermion Green functions 
 shows that Eq. (1) is valid provided the anomalous dimension in the range
$-1<\gamma<1/2$, which corresponds to an enormous range for $\xi$:  $-857<\xi<433$.
Although $\gamma$ is an infrared anomalous dimension and arises from the infrared
behavior of the gauge boson propagator, $\gamma$ itself is not infrared divergent. 

\subsection{Non-zero Temperature}

The question that will be investigated here is the effect of massless photons on
the near mass-shell behavior of the finite-temperature electron  propagator.
At finite temperature the location of the singularity in the propagator is 
temperature-dependent. 
It is convenient to deal with the retarded propagator. 
The inverse of the full retarded thermal propagator is
\begin{equation}
S^{\prime -1}_{R}(P)=\not\!P-m-\Sigma_{R},\label{sinv}\end{equation}
where $\Sigma_{R}$ is defined to contain the  $T=0$ mass counterterm $\delta m$.
In the rest frame of the plasma, rotational invariance requires that
$\Sigma_{R}$ be a linear combination of the matrices $1,\gamma_{0},
\vec{\gamma}\cdot\vec{p}$, and $\gamma_{0}\vec{\gamma}\cdot\vec{p}$. 
It is then straightforward to  compute the inverse of Eq. (\ref{sinv}).
The result may be expressed compactly by defining 
\begin{displaymath}
\tilde{\Sigma}_{R}=\Sigma_{R}-{1\over 2}{\text Tr}\big[\Sigma_{R}\big].
\end{displaymath}
Then ${\text Tr}[\tilde{\Sigma}_{R}]=-{\rm Tr}[\Sigma_{R}]$. The inversion of
Eq. (\ref{sinv}) gives for the full propagator 
\begin{equation}
S^{\prime}_{R}(P)={\not\!P+m+\tilde{\Sigma}_{R}
\over P^{2}-m^{2}-\Pi_{R}(P)}\label{sprime}\end{equation}
where the scalar self-energy in the  denominator is
\begin{equation}
\Pi_{R}(P)={1\over 2}{\rm Tr}\Big[(\not\!P\!+\!m)\Sigma_{R}\Big]
\!-{1\over 4}{\rm Tr}\big[\Sigma_{R}\tilde{\Sigma}_{R}\Big].
\end{equation}
Let $D(P)=P^{2}-m^{2}-\Pi_{R}(P)$ be
 the denominator of Eq. (\ref{sprime}).
The location at which $1/D$ is singular (either a pole or a branch point)
gives a complicated, temperature-dependent relation between $p_{0}$ and $p$
of the form
\begin{equation}
P^{2}=m^{2}+a(P).\label{dispersion}\end{equation}
The  denominator of the propagator has the structure
\begin{equation}
D(P)=\big[P^{2}-m^{2}-a(P)\big]^{1-c(P)}
\big[1+b(P)\big],\end{equation}
with $a,b,c$ generally complex.
 The question of whether the propagator has a  branch point at the
thermal dispersion relation (\ref{dispersion}) is therefore a question of whether
the function $c(P)$ vanishes when  Eq. (\ref{dispersion}) is satisfied.

{\it Order $\alpha$ Approximation for $eT\ll m$:}
In a perturbative expansion the functions $a,b,c$ are each of order $\alpha$ or
smaller. If $eT\ll m$ then $a(P)\ll m^{2}$.
To first order in $\alpha$ the denominator is
\begin{eqnarray}
D(P)\approx 
 P^{2}-m^{2}-&&a(P)
+b(P)(P^{2}-m^{2})\nonumber\\
-&&c(P)(P^{2}-m^{2})\ln(P^{2}-m^{2}).\label{Dapprox}
\end{eqnarray}
Although the   thermal mass-shell 
condition is given by Eq. (\ref{dispersion}), the possibility of a branch point at
the thermal mass shell is reduced to finding whether there is a term of the form
$c(P)(P^{2}-m^{2})\ln(P^{2}-m^{2})$. 
 To first order in $\alpha$ this only requires computing 
\begin{equation}
\Pi_{R}(P)={1\over 2}{\rm Tr}\Big[(\not\!P\!+\!m)\Sigma_{R}\Big]
+{\cal O}(\alpha^{2}).\label{piR}\end{equation}
Note that the logarithmic term in Eq. (\ref{Dapprox}) is 
quite small at the mass-shell (\ref{dispersion}),
${\cal O}(\alpha^{2}\ln(\alpha))$ and was not examined
in previous calculations \cite{nonrelativistic}. 

Section II gives the explict result  for the one-loop electron denominator 
both in Feynman gauge and in general covariant gauges. 
 The detailed calculations are contained in the appendices.

\section{One-Loop Self-Energy}

To compute the one-loop self-energy $\Sigma_{R}$ it will be useful to
use free propagators that are themselves retarded or advanced.
The free retarded propagator for the electron is 
\begin{displaymath}
S_{R}(P)={\not\!P+m\over P^{2}-m^{2}+i\eta p_{0}},\end{displaymath}
and  for
 the photon in a general covariant gauge is
\begin{equation}
D^{\mu\nu}_{R}(K)=\big[\!-g^{\mu\nu}
+(1-\xi){K^{\mu}K^{\nu}\over 2k}{\partial\over\partial k}\big]
{1\over  K^{2}+i\eta k_{0}}.\label{Dmunu}\end{equation}
Since $eT\ll m$ neither propagator requires resummation. 
The advanced propagators are obtained by reversing the
sign of the infinitessimal imaginary part in the denominators.
Dimensional regularization will be used to control the zero-temperature ultraviolet
divergences. The one-loop self-energy has the structure
\begin{equation}
\Sigma_{R}(P)=\Sigma_{R}^{e}(P)+\Sigma_{R}^{\gamma}(P)
+\delta m,\label{deltam}\end{equation}
with $\delta m$  the $T=0$ mass counterterm.
The first contribution has the internal electron on shell:
\begin{eqnarray}
\Sigma_{R}^{e}(P)={ie^{2}\mu^{\epsilon}\over 2}&&\int {d^{4-\epsilon}K
\over (2\pi)^{4}}\,\tanh\big((p_{0}-k_{0})/2T\big)\label{sigmae}\\
\times D^{\mu\nu}_{R}&&(K)\,\gamma_{\mu}
\Big[S_{R}(P-K)-S_{A}(P-K)\Big]\gamma_{\nu}
\nonumber\end{eqnarray}
At $T=0$ this contribution does not contain a term of the form
$(P^{2}-m^{2})\ln(P^{2}-m^{2})$. 
Although the magnitude of the
temperature-dependent part is exponentially suppressed by $\exp(-m/T)$,  that
would  not rule out a branch cut with a small coefficient. 
 Appendix A proves that there is no such term. 

The important contribution is the second term in Eq. (\ref{deltam}). It has the
internal photon on shell:
\begin{eqnarray}
\Sigma_{R}^{\gamma}(P)={ie^{2}\mu^{\epsilon}\over 2}\int {d^{4-\epsilon}K
\over (2\pi)^{4}}&&\coth(k_{0}/2T)\gamma_{\mu}S_{R}(P-K)\gamma_{\nu}\nonumber\\
\times&&\Big[D_{R}^{\mu\nu}(K)-D_{A}^{\mu\nu}(K)\Big]
\nonumber\end{eqnarray}
Inserting the photon propagator (\ref{Dmunu}) gives
\begin{eqnarray}
\Sigma_{R}^{\gamma}(P)={e^{2}\mu^{\epsilon}\over 2}\!\int\!{d^{4-\epsilon}K\over
(2\pi)^{3}} &&\coth(|k_{0}|/2T)\gamma_{\mu}S_{R}(P-K)\gamma_{\nu}
\nonumber\\
\times\big[\!-g^{\mu\nu}&&+{(1-\xi)K^{\mu}K^{\nu}\over 2k}
{\partial \over\partial k}\big]\delta(K^{2}).\label{sigmagamma} 
\end{eqnarray}
The contribution of this term to the denominator of the electron propagator will
be labeled
\begin{equation}
\Pi^{\gamma}(P)={1\over 2}{\rm Tr}\Big[(\not\!P\!+\!m)\Sigma_{R}^{\gamma}(P)\Big]
\label{pie}\end{equation}
Most of the paper is devoted to this computation.

\subsection{Self-Energy in Feynman Gauge}

In the Feynman gauge, $\xi=1$, the photon propagator is
simplest. The trace in Eq. (\ref{pie}) yields 
\begin{equation}
\Pi^{\gamma}(P)=c_{0}+{e^{2} T^{2}\over 6}+f(P)
\end{equation}
where $c_{0}$ is a temperature-independent, divergent constant that is canceled
by the mass counterterm in Eq. (\ref{deltam}).
Thus the mass shell condition (\ref{dispersion}) is essentially
$P^{2}\approx m^{2}+e^{2}T^{2}/6$, which is well known \cite{nonrelativistic}.

The $P$-dependence is contained in the function
\begin{equation}
f(P)={A\mu^{\epsilon}\over 2\pi^{2}}\!\int\!d^{4-\epsilon}K
{\delta(K^{2})\,\coth(|k_{0}|/2T)\over (P_{c}-K)^{2}-m^{2}}\label{f}\end{equation}
\begin{equation}
A=\alpha(P^{2}-3m^{2}),\end{equation}
where $P_{c}=(p_{0}+i\eta,\vec{p})$ because of the retarded prescription. 
The imaginary part of the denominator is $2\eta(p_{0}-k_{0})$ and will generally
change sign as $k_{0}$ is integrated. 
Appendix B  shows that the denominator does not change sign  if $p_{0}$ is
positive time-like:  
\begin{equation}p_{0}>p=|\vec{p}|.\end{equation}
Then $P$ in Eq. (\ref{f}) can be taken real and $m$ replaced by
$m_{c}^{2}=m^{2}-i\eta$. 
The integration is performed in Appendix B with the result
\begin{eqnarray}
f(P)={A\over 2\pi}&&({P^{2}-m^{2}\over P^{2}})\Big[-{1\over \epsilon}
+\ln\big({2\pi T\over \mu}\big)\Big]\label{ftotal}\\
-i{AT\over p}&&\Big[{1\over2}\ln\Big({p_{0}+p\over p_{0}-p}\Big)
+\ln\Big({\Gamma(Z_{+})\over\Gamma(Z_{-})}\Big)\Big].
\nonumber\\
 Z_{\pm}\equiv 1+&&i{m^{2}_{c}-P^{2}\over 4\pi T(p^{0}\pm
p)}.\label{Z}\end{eqnarray}
The ultraviolet divergent term, $1/\epsilon$, is absorbed into the wave
function renormalization factor. 
Various  properties are discussed below.

{\it Analyticity at $P^{2}\approx m^{2}$:} The most important result is that
there is no term of the form $(P^{2}-m^{2})\ln(P^{2}-m^{2})$. In the
vicinity of $P^{2}\approx m^{2}$ the variables $Z_{\pm}$ are close to 1.
Therefore $\ln\Gamma(Z_{\pm})$ is analytic near the mass-shell. This means that
when $T\neq 0$ the  electron propagator has a simple pole at $P^{2}\approx
m^{2}+e^{2}T^{2}/6$ and not a branch cut. 

{\it Zero-Temperature Limit:} It is rather surprising that Eq. (\ref{ftotal}) 
does have a logarithmic branch point precisely at $T=0$. 
This comes about because as $T\to 0$, 
the arguments
$Z_{\pm}\to\infty$ in Eq. (\ref{Z}). Using the Stirling approximation
$\ln\Gamma(Z)\to Z\ln(Z)-Z$ gives the zero-temperature limit 
\begin{eqnarray}
\lim_{T\to 0}{-iT\over p}\ln\Big({\Gamma(Z_{+})\over \Gamma(Z_{-})}\Big)
=&&{P^{2}-m^{2}\over 2\pi
P^{2}}\Big[\ln\Big({im_{c}^{2}-iP^{2}\over 4\pi T\sqrt{P^{2}}}\Big)\nonumber\\
 -&&1+{p_{0}\over 2p}
\ln\Big({p_{0}+p\over p_{0}-p}\Big)\Big].\label{asymp}\end{eqnarray}
Therefore the zero-temperature limit of (\ref{ftotal}) is
\begin{eqnarray}
f(P)\Big|_{T=0}={A(P^{2}-m^{2})\over 2\pi P^{2}}
\Big[-&&{1\over\epsilon}+\ln \big({im_{c}^{2}- iP^{2}\over 2\mu\sqrt{P^{2}}}
\Big)\nonumber\\
-&&1+{p_{0}\over 2p}
\ln\Big({p_{0}+p\over p_{0}-p}\Big)\Big].\end{eqnarray}
This does contain the term $(P^{2}-m^{2})\ln(P^{2}-m^{2})$.
The logarithmic contribution to the electron denominator function,
$D(P)=P^{2}-m^{2}-f(P)$, is
\begin{equation}
D(P)\approx
P^{2}-m^{2}+{\alpha\over\pi}(P^{2}-m^{2})\ln(P^{2}-m^{2}).\label{D}\end{equation}
This agrees  with the standard result in Eq. (\ref{gamma}).

{\it Analyticity for ${\rm Im}\, p_{0}>0$:} A further check of the result is the
requirement that the retarded self-energy 
 be analytic in the upper half of the complex $p_{0}$ plane. 
The only singularities in (\ref{ftotal}) that occur at complex $p_{0}$ come
from poles in
$\Gamma(Z_{\pm})$. 
These occur at $Z_{\pm}=1-n$, for $n$ a positive integer and require
that $p_{0}$ satisfy $p_{0}^{2}=E^{2}-i4\pi nT(p_{0}\pm p)$.
The complex roots of this equation can be written $p_{0}=p_{0r}+ip_{0i}$ and
satisfy
\begin{eqnarray}
p_{0r}^{2}=&&E^{2}+4\pi nTp_{0i}+p_{0i}^{2}\nonumber\\
p_{0i}=&&2\pi nT\big(-1\pm{p\over p_{0r}}\big).\nonumber\end{eqnarray}
If there were a root with $p_{0i}>0$, then the first equation implies that
$|p_{0r}|>E$ so that $p/p_{0r}<1$. But then the second equation implies
that $p_{0i}<0$ contrary to the hypothesis. Hence there are no branch cuts for
${\rm Im}p_{0}>0$.

{\it Imaginary Part:} At $P^{2}=m^{2}$ the self-energy (\ref{ftotal}) is pure
imaginary:
\begin{equation}
f(P)\big|_{P^{2}=m^{2}}=i{\alpha m^{2}T\over p}\ln\big({E+p\over E-p}\big).
\label{imaginary}\end{equation} 
This is an artifact of not having an infrared regularization as shown by
Rebhan in a different context \cite{quark}. Even the sign of Eq.
(\ref{imaginary}) is opposite what it should be for a retared self-energy.  
To check that is an infrared effect, one can return  to Eq.
(\ref{f}) and compute the imaginary part directly:
\begin{eqnarray}
{\rm Im}f(P)={-A\over 2\pi}\!\!\int \!\!{d^{3}k\over
2k}&&\coth\big({k\over 2T}\big)\nonumber\\
\times &&\delta(P^{2}\!-\!m^{2}\!-\!2P\!\cdot\!
K)\Big|_{k_{0}=\pm k}.\nonumber
\end{eqnarray}
At $P^{2}=m^{2}$  the delta function becomes
\begin{displaymath}
\delta(2P\cdot K)\big|_{k_{0}=\pm k}={\delta(k)\over 2(E\mp \vec{p}\cdot\hat{k})}
.\end{displaymath}
Even though the support is at $k=0$ the integral does not vanish because
because of the Bose-Einstein enhancement of $k=0$:
\begin{displaymath}
\int_{0}^{\infty}k\, dk\coth\big({k\over 2T}\big)\delta(k)=T.\end{displaymath} 
The remaining angular integration is
\begin{displaymath}
{\rm Im} f(P)={\alpha m^{2}TE\over 2\pi}\int d\Omega {1\over
E^{2}-(\vec{p}\cdot\hat{k})^{2}}\end{displaymath}
and reproduces Eq. (\ref{imaginary}). The entire effect comes from the point
$k=0$. If the infrared behavior is regulated there will be no imaginary part at
$P^{2}=m^{2}$.

\subsection{Self-Energy in General Covariant Gauge}

In a general covariant gauge with $\xi\neq 1$, the second term in Eq.
(\ref{sigmagamma}) must be computed. A prime will be used to denote this
contribution:
\begin{eqnarray}
\Sigma_{R}^{\prime}(P)=(1-\xi){e^{2}\mu^{\epsilon}\over
2}\!\int\!&&{d^{4-\epsilon}K\over (2\pi)^{3}}{\partial \delta(K^{2})\over\partial
k}
\coth\big({|k_{0}|\over 2T}\big)\nonumber\\
\times&&{K^{\mu}K^{\nu}\over
2k}\gamma_{\mu}S_{R}(P-K)\gamma_{\nu}\end{eqnarray}
The trace necessary for the electron  denominator is
\begin{equation}
f^{\prime}(P)={1\over 2}{\rm
Tr}\big[(\not\!P+m)\Sigma_{R}^{\prime}(P)\big],\end{equation} which yields
\begin{eqnarray}
f^{\prime}(P)={B\mu^{\epsilon}\over 4\pi^{2}}&& \int {d^{4-\epsilon}K\over k}
{\partial \delta(K^{2})\over \partial k}\nonumber\\
&&\coth({|k_{0}|\over 2T})
{P\cdot K-K^{2}\over (P_{c}-K)^{2}-m^{2}}\label{fprime}\end{eqnarray}
\begin{equation}
B=\alpha(1-\xi)(P^{2}-m^{2})\end{equation}
In Appendix C this is computed for $p_{0}>p$ with the result
\begin{eqnarray}f^{\prime}(P)={B\over 2\pi}&&\Big[{1\over\epsilon}-1
+\ln\big({\mu\over 2\pi T}\big)
+{i2\pi p_{0}T\over P^{2}-m_{c}^{2}}\nonumber\\
+&&{i\pi T\over 2p}\ln\Big({p_{0}+p\over p_{0}-p}\Big)+{i\pi T\over p}
\ln\Big({\Gamma(Z_{+})\over\Gamma(Z_{-})}\Big)\nonumber\\
-&&{1\over 2}(1+{P^{2}-m^{2}\over 2p(p_{0}+p)})\psi(Z_{+})\nonumber\\
-&&{1\over 2}(1-{P^{2}-m^{2}\over 2p(p_{0}-p)})\psi(Z_{-})
\Big]\label{ftotalprime}\end{eqnarray}

{\it Analyticity at $P^{2}\approx m^{2}$:} As was the case in Feynman gauge, there
is no explicit $\ln(im^{2}-iP^{2})$. Since $\Gamma(Z_{\pm})$ and $\psi(Z_{\pm})$
are analytic near $Z_{\pm}\approx 1$, the entire function $f^{\prime}(P)$ is
analytic near the mass shell. There is no branch point.

{\it Zero-Temperature Limit:} To evaluate $f^{\prime}(P)$ at zero temperature
requires using  Eq. (\ref{asymp}) and the asymptotic behavior
$\psi(Z_{\pm})\to\ln(Z_{\pm})$ in order  to 
obtain
\begin{eqnarray}
 f^{\prime}(P)\Big|_{T=0}={B\over 2\pi}\Big[{1\over
\epsilon}-{P^{2}+m^{2}\over 2P^{2}} +\ln\big({2\mu\sqrt{P^{2}}\over
im^{2}-iP^{2}}\big)\Big].
\end{eqnarray}
This does contain the  term $(P^{2}-m^{2})\ln(P^{2}-m^{2})$ with coefficient 
\begin{displaymath}
-{\alpha\over 2\pi}(1-\xi)(P^{2}-m^{2})\ln(P^{2}-m^{2}).\end{displaymath}
Subtracting  this from the Feynman-gauge contribution (\ref{D}) gives
\begin{equation}
D(P)\!\approx\! P^{2}\!-\!m^{2}+{\alpha\over
2\pi}(3\!-\xi)(P^{2}\!-m^{2})\ln(P^{2}\!-m^{2}).
\end{equation}
This agrees with the general result (\ref{gamma}).

{\it Analyticity for Im $p_{0}>0$:} Since the only singularities in
$\Gamma(Z_{pm})$ or $\psi(Z_{\pm})$ are when $Z_{\pm}$ is zero or a negative
integer, the same analysis as before shows that (\ref{ftotalprime}) is analytic
in the upper-half of the complex $p_{0}$ plane. 

{\it Imaginary Part:} The factor $B$ vanishes at $P^{2}=m^{2}$.
However, from the first line of Eq. (\ref{ftotalprime}) it appears that
$f^{\prime}(P)\to i\alpha(1-\xi) ET$ as $P^{2}\to m^{2}$,
As before this term would not survive  if the   infrared behavior had
been regulated.

\section{Comments}

The branch point in the $T=0$ electron propagator is a major
complication. 
It indicates the impossibility  an   electron being isolated. It will
always have a cloud of photons and will therefore not be 
  an eigenstate of the mass operator \cite{Schroer}. To treat charged
particles properly it is necessary to employ a Hilbert space containing an
infinite number of coherent photons \cite{Chung,Kibble,Kulish}. 
The LSZ 
asymptotic conditions and  reduction formulas  are modified \cite{Zwanziger}.

It is remarkable that for $0<eT\ll m$  the electron self-energy does not 
contain a term $(P^{2}-m^{2})\ln(P^{2}-m^{2})$.
The electron propagator thus has a
simple pole at the thermal mass shell $P^{2}\approx m^{2}+e^{2}T^{2}/6$.   
 This result does not automtically carry over to QCD at low
temperature.  No matter how  large the quark masses are that 
break chiral symmetry, the  gluon propagator  requires 
 Braaten-Pisarski resummation.

\acknowledgements

This work was supported in part by the U.S. National Science Foundation under
grant PHY-9630149.

\appendix
\section{Analysis of $\Sigma_{R}^{\lowercase{e}}$}

One can eliminate the self-energy contribution in Eq. (\ref{sigmae}) as a
possibility for  producing a branch cut at $P^{2}=m^{2}$ rather easily.
\begin{eqnarray}
\Sigma_{R}^{e}(P)={e^{2}\mu^{\epsilon}\over 2}&&\!\int\!{d^{4-\epsilon}K\over
(2\pi)^{3}}
\delta[(P-K)^{2}-m^{2}]\nonumber\\
\times&&\tanh({|p_{0}-k_{0}|\over
2T})D_{R}^{\mu\nu}(K)\gamma_{\mu}(\not\!P-\not\!K+m)\gamma_{\nu}
\nonumber\end{eqnarray}
The delta function constraint  sets
$k_{0}=p_{0}\pm \Omega$ where $\Omega=(m^{2}+(\vec{p}-\vec{k})^{2})^{1/2}$
so that
\begin{eqnarray}\Sigma_{R}^{e}(P)
={e^{2}\mu^{\epsilon}\over 16\pi^{3}}\int &&{d^{3-\epsilon}k\over
2\Omega}\tanh({\Omega\over 2T})\nonumber\\
\times&&D_{R}^{\mu\nu}(K)\gamma_{\mu}(\not\!P-\not\!K+m)\gamma_{\nu}
\Big|_{k_{0}=p_{0}\pm\Omega}\nonumber
\end{eqnarray}
If this were to contain a term $(P^{2}-m^{2})\ln(P^{2}-m^{2})$ then the
derivative with respect to $p_{0}$ would be logarithmically divergent at
$p_{0}=E$.  The case
$k_{0}=p_{0}+\Omega$ does not have this behavior because  as
$p_{0}\to E$
 the denominator of the photon propagator is infrared safe. 
That leaves the case $k_{0}=p_{0}-\Omega$. 
The largest contribution to
the derivative of the self-energy comes from differentiating $1/K^{2}$:
\begin{displaymath}
{\partial\Sigma_{R}^{e}(P)\over\partial p_{0}}\Big|_{p_{0}=E}
\sim \int {d^{3-\epsilon}k\over 2\Omega}\;{E-\Omega\over
[(E-\Omega)^{2}-k^{2}]^{2}}.
\end{displaymath}
The important region is $k$ small, in which case 
 $\Omega\approx E-\vec{v}\cdot\vec{k}$,
where $\vec{v}=\vec{p}/E$ is the electron velocity. 
This gives
\begin{displaymath}
\int{d^{3-\epsilon}k\over 2E}\;{\vec{v}\cdot\vec{k}+{\cal O}(k^{2})
\over [(\vec{v}\cdot\vec{k})^{2}-k^{2}]^{2}}.\end{displaymath}
By power counting this integration could give a logarithmic divergence. However
 the numerator $\vec{v}\cdot\vec{k}$ is odd in $\vec{k}$ and thus the angular
integral gives zero. The neglected terms are all $d^{3}k\, k^{2}/k^{4}$ and are
 finite. Thus $\Sigma_{R}^{e}$ cannot contain a term
$(P^{2}-m^{2})\ln(P^{2}-m^{2})$.

\section{Calculation of $\Sigma_{R}^{\gamma}$ in Feynman gauge}

The integral displayed in Eq. (\ref{f}) for the Feynman-gauge self-energy is
performed explicitly in this Appendix. The answer for the zero-temperature
contribution is displayed in Eq. (\ref{f0}); for the thermal contribution, in Eq.
(\ref{fT}). The sum of the two gives the result quoted in Eq. (\ref{ftotal}).

To analyze Eq. (\ref{f}) the integration 
 over $k_{0}$ and over angles can be performed with the result
\begin{displaymath}
f(P)={A\mu^{\epsilon}\over 4\pi
p}\int_{0}^{\infty}{dk\over k^{\epsilon}}\ln\Big[{(k+r)(k-s)\over
(k-r)(k+s)}\Big]\coth({k\over 2T})\end{displaymath}
\begin{equation}
r={P_{c}^{2}-m^{2}\over 2(p_{c}^{0}+p)}
\hskip1cm s={P_{c}^{2}-m^{2}\over 2(p_{c}^{0}-p)},\end{equation}
where $p_{c}^{0}=p^{0}+i\eta$. The imaginary parts of $r$ and $s$ are always
positive for any real values of $p_{0}$ and $p$. For $p_{0}$ positive and
time-like, i.e. $p_{0}>p$, the denominators of (B1) are positive  so that
$r$ and $s$ can be replaced by
\begin{displaymath}
r'={P^{2}-m^{2}+i\eta\over2 (p_{0}+p)}
\hskip1cm
s'={P^{2}-m^{2}+i\eta\over2 (p_{0}-p)}.\end{displaymath}
 This is the same as using a Feynman prescription, $m^{2}-i\eta$, in the original
denominator of (\ref{f}). Thus
\begin{equation}
f(P)={A\mu^{\epsilon}\over 2\pi^{2}}\!\int\!d^{4-\epsilon}K
{\delta(K^{2})\,\coth(|k_{0}|/2T)\over P^{2}-m^{2}-2P\cdot
K+i\eta},\label{b2}\end{equation}
which will be easier to compute.
Because the imaginary part of the denominator no longer changes sign
one can  use   the parametric
representation
\begin{equation}{1\over X+i\eta}=-i\int_{0}^{\infty}ds\;e^{i(X+i\eta)s}
\label{identity},\end{equation}
and interchange the order of integrations to get
\begin{eqnarray}f(P)=&&-i\int_{0}^{\infty}ds\;e^{i(P^{2}-m^{2}+i\eta)s}J(s)
\nonumber\\
J(s)\equiv&&{A\mu^{\epsilon}\over 2\pi^{2}}\int d^{4-\epsilon}K
\delta(K^{2})\coth(|k^{0}|/2T)\;e^{-i2P\cdot K s}\nonumber.\end{eqnarray}
The $k^{0}$ integration can be performed using the Dirac delta function and the angular integrals
are elementary:
\begin{eqnarray}J(s)={A\mu^{\epsilon}\over 2\pi ps}
&&\int_{0}^{\infty}\!dk\;k^{-\epsilon}
\coth(k/2T)\label{J}\\
\times&&\big(\sin[2s(p^{0}+p)k]-\sin[2s(p^{0}-p)k]\big).\nonumber
\end{eqnarray}
At zero temperature there are ultraviolet divergences (regularized by
$k^{-\epsilon}$); at finite temperature there are not.
The relation $\coth(k/2T)=1+2n(k/T)$, where
\begin{equation}n(x)={1\over \exp(x)-1},\label{bose}\end{equation}
allows the temperature-independent
part of the integration to be isolated and leads to
\begin{displaymath}J(s)=J_{0}(s)+J_{T}(s).\end{displaymath} 

{\it Zero-Temperature:} At $T=0$ the momentum integration in Eq. (\ref{J})
 gives
\begin{eqnarray}&&J_{0}(s)=Ns^{\epsilon-2}\nonumber\\
N\equiv{A (2\mu)^{\epsilon}\over 4\pi
p}\Gamma(1-\epsilon)&&\cos({\epsilon\pi\over 2})
\bigl[(p^{0}+p)^{\epsilon-1}-(p^{0}-p)^{\epsilon-1}\bigr].\nonumber\end{eqnarray}
The zero-temperature contribution to $f(P)$ is 
\begin{eqnarray}f_{0}(P)=-iN&&\int_{0}^{\infty}ds\;e^{i(P^{2}-m^{2}+i\eta)s}
s^{\epsilon-2}\nonumber\\
=-iN&&\Gamma(\epsilon-1)(im^{2}-iP^{2})^{1-\epsilon}.\nonumber\end{eqnarray}
In the limit $\epsilon\to 0$ this has the expected $1/\epsilon$ ultraviolet divergence plus finite
terms:
\begin{eqnarray}f_{0}(P)={A(m^{2}-P^{2})\over 2\pi P^{2}}
\Big[ {1\over\epsilon}+1-&&{p^{0}\over 2p}\ln({p^{0}+p\over p^{0}-p})
\nonumber\\
+&&\ln\Big({2\mu\sqrt{P^{2}}\over
im^{2}-iP^{2}}\Big)\Big].\label{f0}\end{eqnarray} 
 There is no branch point at
$p_{0}=p$  because of a cancellation between the two logarithms. It does have a
logarithmic branch cut $P^{2}=m^{2}$ as expected.

{\it  Thermal Contribution:} The remainder
 of   Eq. (\ref{J}) is temperature-dependent:
 \begin{displaymath}J_{T}(s)\!=\!{A\mu^{\epsilon}\over
\pi ps}\!\!\int_{0}^{\infty}\!\!dk k^{-\epsilon}
{\sin[2s(p^{0}\!+p)k]\!-\sin[2s(p^{0}\!-p)k]\over\exp(k/T)-1}.
\end{displaymath}
This can be performed using the useful integral \cite{Gradshteyn}
\begin{equation}
\int_{0}^{\infty}\!dx\, x^{\nu-1}{\exp(a x)\over
\exp(bx)-1} ={\Gamma(\nu)\over b^{\nu}}\zeta[\nu,1-{a\over b}],
\label{integral}\end{equation}
which is valid for positive real $b$ and  ${\rm Re}\, a <  b$ . 
The result is
\begin{eqnarray}J_{T}(s)={A T\over ps}
\Big[&& {-1\over 4\pi T(p^{0}+p)s}+n\big(4\pi T(p^{0}+p)s\big)\nonumber\\
+&&{1\over 4\pi T(p^{0}-p)s}-n\big(4\pi
T(p^{0}-p)s\big)\Big]\nonumber\end{eqnarray} The final integration over $s$ 
requires
\begin{displaymath}f_{T}(P)=-i\int_{0}^{\infty}ds\;e^{i(P^{2}-m^{2}+i\eta)s}
J_{T}(s).\end{displaymath}
Although various pieces of $J_{T}(s)$ behave like $s^{-2}$ for small $s$, the
complete function is completely finite at $s=0$. To
integrate over
$s$ it is convenient  to regulate the
small $s$  behavior of the individual terms by  multiplying the integrand by
$s^{\nu}$ with $\nu>1$.  The terms $s^{\nu-2}$ 
 integrate to  Gamma functions. The  exponential parts can
be integrated by using (\ref{integral}) again. The full integration has no
singularity at $\nu=1$ or at $\nu=0$.
 After setting $\nu\to 0$ the  result is
\begin{eqnarray}f_{T}(P)=&&{A(m^{2}-P^{2})\over2\pi P^{2}}
\Big[-1+{p_{0}\over 2p}\ln({p^{0}+p\over p^{0}-p})\nonumber\\
&&\hskip2cm + \ln\big({im^{2}-iP^{2}\over 4\pi T\sqrt{P^{2}}}\big)\Big]\label{fT}\\
-&&{iA T\over p}\Big[{1\over 2}\ln({p^{0}+p\over p^{0}-p})
+\ln\Big({\Gamma(Z_{+})
\over \Gamma(Z_{-})}\Big)\Big],\nonumber\end{eqnarray}
where the arguments of the gamma function are
\begin{equation}
 Z_{\pm}\equiv 1+i{m^{2}-P^{2}\over 4\pi T(p^{0}\pm
p)}.\label{Z2}\end{equation}
 Although it is not apparent, $f_{T}(P)$ does vanishes $T=0$.
The most important feature is the $(m^{2}-P^{2})\ln(im^{2}-iP^{2})$ term that
exactly cancels the zero-temperature contribution (\ref{f0}). The sum of Eqs.
(\ref{f0}) and (\ref{fT}) is given  in Eq. (\ref{ftotal}). 

\section{Calculation of $\Sigma_{R}^{\prime}(P)$ in Covariant gauge}

This Appendix computes the self-energy integral (\ref{fprime}),
which is present in  covariant gauges in which  $\xi\neq 1$. The
$T=0$ result is displayed in Eq. (\ref{f0prime}) and the temperature-dependent
part in Eq. (\ref{fTprime}). 

The analysis  begins with the observation that the denominators of (\ref{fprime})
have singularities in $k$ at the locations (B1). Therefore for $p^{0}>p$  the
infinitessimal positive imaginary part can be omitted from
$p_{c}^{0}$ and replaced by a negative imaginary part on the mass: $m^{2}\to
m_{c}^{2}=m^{2}-i\eta$ as was done in Eq. (B2). After an integration by parts, 
Eq. (\ref{fprime}) can be written
\begin{eqnarray}
f^{\prime}(P)={B\mu^{\epsilon}\over 4\pi^{2}}&& \int {d^{4-\epsilon}K\over k^{2}}
\delta(K^{2})\coth(|k_{0}|/ 2T)\nonumber\\
&&[1\!-\epsilon+k{\partial\over\partial k}]{-P\cdot K+K^{2}\over
(P-K)^{2}-m_{c}^{2}}.\end{eqnarray}
It is convenient to put $\sigma=1-\epsilon$. 
Computing the derivatives gives
\begin{eqnarray}
f^{\prime}(P)=&&{B\mu^{\epsilon}\over 4\pi^{2}} \int {d^{4-\epsilon}K\over k^{2}}
\delta(K^{2})\coth(|k_{0}|/2T)\nonumber\\
\times &&\Big[{-\sigma P\cdot K+\vec{p}\cdot \vec{k}-2k^{2}\over
(P-K)^{2}-m_{c}^{2}} +{2P\cdot K(\vec{p}\cdot\vec{k}-k^{2})
\over ((P-K)^{2}-m_{c}^{2})^{2}}\Big].\nonumber\end{eqnarray}
 The denominators can be exponentiated using Eq. (\ref{identity}) so that
\begin{eqnarray}
f^{\prime}(P)=&&-i\int_{0}^{\infty}ds\,e^{i(P^{2}-m^{2}+i\eta)s}
J^{\prime}(s)\nonumber\\
J^{\prime}(s)\equiv &&{B\mu^{\epsilon}\over 4\pi^{2}} \int {d^{4-\epsilon}K\over
k^{2}}
\delta(K^{2})\coth(|k_{0}|/2T)\,e^{-i2P\cdot K s}\nonumber\\
\times &&\Big[\!-\sigma P\cdot K+\vec{p}\cdot\vec{k}-2k^{2}\! 
-is 2P\cdot K(\vec{p}\cdot\vec{k}-k^{2})
\Big].\nonumber
\end{eqnarray}
Integration over $k_{0}$ and the angles of $\vec{k}$ give
\begin{eqnarray}
J^{\prime}(s)=&&{B\mu^{\epsilon}\over 4\pi p}\int_{0}^{\infty}\!dk
\,k^{-\epsilon}\coth(k/2T)\label{jprime}
\\
\times\Big[&&[2ip(p_{0}\!+p)-{1\over s}D_{+}
-{1\over s^{2}}{i\epsilon D_{-}\over
2k^{2}}]\sin[2sk(p_{0}\!+p)]\nonumber\\ +&&[2ip(p_{0}\!-p)+{1\over s}D_{+}-{1\over
s^{2}}{i\epsilon D_{-}\over 2k^{2}}]\sin[2sk(p_{0}\!-p)]\Big]\nonumber
\end{eqnarray}
where $D_{\pm}\equiv 1\pm s\partial/\partial s$.
Note that the integral is convergent at both small and large $k$. The term linear
in $\epsilon$ cannot be omitted as it will lead to a nonvanishing contribution
after the $s$ integration. As
before use
$\coth(k/2T)=1+2n(k/T)$ to obtain the
separation
\begin{displaymath}
J^{\prime}(s)=J^{\prime}_{0}(s)+J^{\prime}_{T}(s).\end{displaymath}
    
{\it Zero-Temperature:} The zero-temperature integration
of (\ref{jprime}) contains powers of
$k$ times a sin function. The result is
\begin{eqnarray}
J_{0}^{\prime}(s)=&&{B\mu^{\epsilon}\over 4\pi }\Gamma(1-\epsilon)
\cos({\epsilon\pi\over 2})\nonumber\\
\times&&{[2(p_{0}+p)]^{\epsilon}\over p}\big[{-\epsilon s^{\epsilon-2}\over 2
(p_{0}+p)}+i[p-\epsilon(p_{0}+p)] s^{\epsilon-1}\big]\nonumber\\
+&&(p\to -p).\nonumber\end{eqnarray}
Because only powers of $s$ appear, the final integration over $s$
\begin{displaymath}
f^{\prime}_{0}(P)=-i\int_{0}^{\infty}ds\,e^{i(P^{2}-m^{2}+i\eta)s}\,J^{\prime}_{0}(s)
\end{displaymath} is straightforward. The result in the limit $\epsilon\to 0$ is  
\begin{eqnarray}
f^{\prime}_{0}(P)={B\over 2\pi}\Big[{1\over \epsilon}-{P^{2}+m^{2}\over 2P^{2}}
+\ln\big({2\mu\sqrt{P^{2}}\over im^{2}-iP^{2}}\big)\Big].
\label{f0prime}\end{eqnarray}

{\it Thermal Contribution:}
Since there are no ultraviolet divergences in  the thermal part of Eq.
(\ref{jprime}) one may set  $\epsilon=0$:
\begin{eqnarray}
J^{\prime}_{T}(s)=&&{B\over 2\pi p}\int_{0}^{\infty}\!dk
\,n(k/T)\nonumber\\
\times\Big[&&[2ip(p_{0}+p)-{1\over s}D_{+}]\sin[2sk(p_{0}+p)]\nonumber\\
+&&[2ip(p_{0}-p)+{1\over s}D_{+}]\sin[2sk(p_{0}-p)]\Big].\nonumber
\end{eqnarray}
 This can be performed using Eq. (\ref{integral}) and gives
\begin{eqnarray}
J^{\prime}_{T}(s)=&&B({-i\over 2\pi s}+ip_{0}T)\nonumber\\
+B&&T[i(p_{0}+p)-{1\over 2ps}(1+T{\partial\over\partial T})]\,n\big(4\pi
T(p_{0}+p)s\big)\nonumber\\ 
+B&&T[i(p_{0}-p)+{1\over 2ps}(1+T{\partial \over \partial T})]\,
n\big(4\pi T(p_{0}-p)s\big).
\nonumber\end{eqnarray}
For later convenience the $s$ derivatives have been converted to $T$ derivatives
using
\begin{displaymath}s{\partial\over\partial s}n(aTs)
=T{\partial\over\partial
T}n(aTs).\end{displaymath} The remaining integration is
\begin{displaymath}
J^{\prime}_{T}(P)=-i\int_{0}^{\infty}ds\,e^{i(P^{2}-m^{2}+i\eta)s}
\,J_{T}^{\prime}(s)\end{displaymath}
It is easy to check that $J_{T}^{\prime}(s)$ vanishes at $s=0$. However, various
pieces behave as $s^{-2}$ and $s^{-1}$ at small $s$. It is therefore convenient
to multiply $J_{T}(s)$ by a factor $s^{\nu}$ where initially $\nu>1$. Individual
terms will have singularities at $\nu=1$ and at $\nu=0$. All these singularities
cancel when the terms are combined, at which point one can put $\nu=0$. The
integration is performed using Eq. (\ref{integral}). The $T$ derivatives of
Euler gamma functions give $\psi$ functions.
The final result is
\begin{eqnarray}
f_{T}^{\prime}(P)={B\over 2\pi}&&\Big[\ln\Big({im^{2}-iP^{2}\over 4\pi
T\sqrt{P^{2}}}\Big)
-{P^{2}-m^{2}\over 2P^{2}}+{i2\pi p_{0}T\over P^{2}-m_{c}^{2}}\nonumber\\
+&&{i\pi T\over 2p}\ln\Big({p_{0}+p\over p_{0}-p}\Big)+{i\pi T\over p}
\ln\Big({\Gamma(Z_{+})\over\Gamma(Z_{-})}\Big)\nonumber\\
-&&{1\over 2}(1+{P^{2}-m^{2}\over 2p(p_{0}+p)})\psi(Z_{+})\nonumber\\
-&&{1\over 2}(1-{P^{2}-m^{2}\over 2p(p_{0}-p)})\psi(Z_{-})
\Big]\label{fTprime}\end{eqnarray}
with $Z_{\pm}$ as in (\ref{Z2}).
The sum of Eqs. (\ref{f0prime}) and (\ref{fTprime}) is displayed in
Eq. (\ref{ftotalprime}).

\end{document}